\documentstyle[12pt]{article}

\makeatletter
\renewcommand\thesubsection{\thesection.\@arabic\c@subsection}
\makeatother

\newcommand{\sect}[1]{\setcounter{equation}{0}\section{#1}}

\newcommand {\beq}{\begin{equation}}
\newcommand {\eeq}{\end{equation}}
\newcommand {\beqa}{\begin{eqnarray}}
\newcommand {\eeqa}{\end{eqnarray}}         
\newcommand {\beqs}{\begin{eqnarray*}}
\newcommand {\eeqs}{\end{eqnarray*}}
\newcommand {\bds}{\begin{displaymath}}
\newcommand {\eds}{\end{displaymath}}
\newcommand {\n}{\nonumber\\}

\newcommand {\bebb}{\begin{thebibliography}}
\newcommand {\eebb}{\end{thebibliography}}      
\newcommand {\bbit}{\bibitem}

\def\pd{\prod}

\def\ra{\rangle}

\def\dg{\dagger}

\def\ph{\phi}

\def\journal#1&#2(#3){\unskip, \sl #1\ \bf #2 \rm(19#3) }
\def\andjournal#1&#2(#3){\sl #1~\bf #2 \rm (19#3) }

\def\dz{\frac{d}{dz}}

\begin{document}


\begin{flushright}
\end{flushright}

\vskip 1cm

\begin{center}
{\Large\bf Polynomial algebras and exact solutions of general quantum non-linear optical models I:
Two-mode boson systems}

\vspace{1cm}

{\large Yuan-Harng Lee $^a$, Wen-Li Yang $^{b}$ and Yao-Zhong Zhang $^{a}$}
\vskip.1in

$a.$ {\em School of Mathematics and Physics,
The University of Queensland, Brisbane, Qld 4072, Australia}

$b.$ {\em Institute of Modern Physics, Northwest University,
Xi'an 710069, China}

\end{center}

\date{}



\begin{abstract}
We introduce higher order polynomial deformations of $A_1$ Lie algebra.
We construct their unitary representations and the corresponding single-variable
differential operator realizations. We then use the results to obtain exact
(Bethe ansatz) solutions to a class of 2-mode boson systems, including the
Boson-Einstein Condensate models as special cases. Up to an overall factor,
the eigenfunctions of the 2-mode boson systems are given by polynomials whose roots are
solutions of the associated Bethe ansatz equations. The corresponding eigenvalues
are expressed in terms of these roots. We also establish the spectral equivalence
between the BEC models and certain quasi-exactly solvable Sch\"ordinger potentials.

\end{abstract}

\vskip.1in





\setcounter{section}{0}
\setcounter{equation}{0}
\sect{Introduction}

Polynomial algebras are non-linear deformations of Lie algebras
and have recently found widespread applications in theoretical physics whereby
they appear in diverse topics such as quantum mechanics, Yang-Mills type gauge theories,  quantum non-linear optics,
integrable systems and (quasi-)exactly solvable models, to name a few (see e.g. \cite{Higgs79}-\cite{Klishevich01}).

One of the reasons for their increasing prevalence stems from the realization that traditional linear
Lie algebras describe only a very restrictive subset of linear symmetries and that many physical systems
do in fact possess non-linear symmetries such as those
in which the commutations of the symmetry algebra generators return polynomial terms.

Due to their importance, a number of studies have been undertaken to investigate the mathematical properties
of these algebras \cite{Smith90}. In particular, differential realizations of certain quadratic and cubic algebras
have been explored in \cite{Zhedanov92,Kumar01}
and also in \cite{Beckers99,Debergh00} in connection with the theory of quasi-exact integrability
\cite{Turbiner88,Ushveridze94,Gonzarez93}.

In this article, we introduce a novel class of higher order polynomial deformations of the classical $A_1$
Lie algebra and construct their unitary representations in terms of boson operators and single-variable
differential operators. We will then use the differential realizations of these algebras and the
Functional Bethe Ansatz method (see e.g. \cite{Wiegmann94,Sasaki09}) to obtain one of the main results of this paper,
that is the exact eigenfunctions and energy
eigenvalues of the following class of Hamiltonians
\beq
H=\sum_i^2 w_iN_i+\sum_{i,j}^2 w_{ij}N_iN_j + g\left( a_1^{\dagger s}a_2^r +a_1^sa_2^{\dagger r} \right),
\hspace{1cm} r,s \in  {\bf N} \label{bosonH1},
\eeq
where and throughout $a_i~(a_i^\dagger)$ are boson or photon annihilation (creation) operators with frequencies $w_i$, 
$N_i=a^\dagger_ia_i$ are number operators, $w_{ij}$ and $g$ are real coupling constants. Without loss
of generality, in the following we will identify $w_{12}$ with $w_{21}$.
Hamiltonians (\ref{bosonH1}) appear in the description of various physical systems of interest
such as non-linear optics \cite{Karassiov94,Karassiov02} and Bose-Einstein Condensates (BECs)
\cite{Anderson95}-\cite{Donley02}. For instance, the non-diagonal terms in (\ref{bosonH1}) describe processes
of multi-photon scattering and higher-order harmonic generation in quantum nonlinear optics.
Let us point out that (\ref{bosonH1}) is a two-mode version of the more general multi-mode Hamiltonian considered in
\cite{Alvarez02} in which the quasi-exact solvability of the multi-mode system was established 
by a different procedure and without giving exact solutions (see also \cite{Alvarez95} for the case of third-order
harmonic generation)\footnote{We became aware of these three references after submitting our work. We thank
one referee for pointing them out.}. Hamiltonians for the
special cases of $s=r=1$ and $s=2,~r=1$ have also been studied using Algebraic Bethe Ansatz (ABA) method \cite{Links03}.

This paper is organized as follows, in section 2 we propose a class of generalized polynomial $su(1,1)$
algebras and derive their boson realizations. In section 3, we use these deformed $su(1,1)$ algebras
as base elements to generate higher order polynomial algebras via a Jordan-Schwinger like construction method.
We then identify these algebras as the dynamical algebra of the Hamiltonian (\ref{bosonH1}) in section 4
and solve for the eigenvalue problem in general via the Functional Bethe Ansatz method. In section 5,
we present explicit results for the Hamiltonian (\ref{bosonH1}) when $r,s \le 2$ and $r=s=3$. In section 6,
we establish the spectral correspondence of these specific models with quasi-exactly solvable (QES)
Schr\"odinger potentials. Finally we summarize our results in section 7 and discuss further avenues of investigation.



\sect{Polynomial deformations of su(1,1) algebra}

Let $k$ be a positive integer, $k=1,2,\cdots$. We start off by proposing a class of polynomial algebras
of degree $k-1$ defined by the commutation relations
\beqa
[ Q_0,Q_{\pm} ] &=& \pm Q_{\pm}, \n
\left[ Q_+,Q_{-} \right]  &=& \ph^{(k)}(Q_0)-\ph^{(k)}(Q_0-1), \label{su11-poly-alg}
\eeqa
where
\beq
\ph^{(k)}(Q_0)= - \pd_{i=1}^{k} \left( Q_0+\frac{i}{k}-\frac{1}{k^2} \right)
+\pd_{i=1}^k \left(\frac{i-k}{k} - \frac{1}{k^2} \right)
\eeq
is a $k^{th}$-order polynomial in $Q_0$. The algebra admits Casimir operator of the following form
\beq
C=Q_-Q_+ +\ph^{(k)}(Q_0)=Q_+Q_-+\ph^{(k)}(Q_0-1).\label{casimir}
\eeq
For $k=1$ and $k=2$, (\ref{su11-poly-alg}) reduces to the oscillator and $su(1,1)$ algebras, respectively.
Thus, the algebra (\ref{su11-poly-alg}) can be viewed as polynomial extensions of the linear $su(1,1)$
and oscillator algebras.

Similar to the $su(1,1)$ algebra case, unitary representations of (\ref{su11-poly-alg}) are infinite dimensional.
In this section, we shall concentrate on the following one-mode boson realization of the algebra,
\beq
Q_+=\frac{1}{(\sqrt{k})^k}(a^{\dg})^k ,~~~~~
Q_-=\frac{1}{(\sqrt{k})^k}(a)^k , ~~~~~
Q_0 =\frac{1}{k} \left(a^{\dg}a+ \frac{1}{k} \right).\label{su11-poly-boson}
\eeq
In this realization, the Casimir (\ref{casimir}) takes the particular value,
\beq
C=\pd_{i=1}^k \left(\frac{i-k}{k} - \frac{1}{k^2} \right).\label{casimir-value}
\eeq

We now construct the unitary representations corresponding to the realization (\ref{su11-poly-boson})
in the Fock space ${\cal H}_b $. There are $k$ lowest weight states,
\beqa
|0\ra , \hspace{0.2cm} (a^\dg)|0\ra  \hspace{0.2cm} , ..., \hspace{0.2cm} (a^\dg)^{k-1}|0\ra.\label{lw-state1}
\eeqa
Writing these lowest weight states as $|q,0\ra$ using the Bargmann index $q$, we have
\beq
Q_0|q,0\ra=q|q,0\ra, ~~~~~Q_-|q,0 \ra =0.
\eeq
It follows from (\ref{casimir}) and (\ref{casimir-value}) that
$\pd_{i=1}^{k}\left(q+\frac{i-k}{k}-\frac{1}{k^2}\right) =0$,
from which we get
\beq
q= \frac{1}{k^2} ,~ \frac{k+1}{k^2},~\frac{2k+1}{k^2} ,\cdots, ~\frac{(k-1)k+1}{k^2}.\label{q-values}
\eeq
This means that the boson realization (\ref{su11-poly-boson}) corresponds to the infinite dimensional
unitary representation with particular $ q $ values (\ref{q-values}). In other words, the $ {\cal {H}} _b $
decomposes into the direct sum
$ {\cal H}_b = {\cal H}_{b}^{\frac{1}{k^2}}\oplus \cdots\oplus {\cal H}_b^{\frac{(k-1)k+1}{k^2}}$
of $k$ irreducible components ${\cal H}_b^{\frac{1}{k^2}},..., {\cal H}_b^{\frac{(k-1)k+1}{k^2}}$.

Noting $kq-\frac{1}{k}=0,1,..,k-1$ for all the allowed $q$ values given in (\ref{q-values}),
we can write the lowest weight states (\ref{lw-state1}) as
$|q,0 \ra=(a^{\dg})^{kq-\frac{1}{k}}|0 \ra$. 
The general Fock states $|q,n \ra \sim Q_+^n|q,0 \ra $ in the irreducible representation space ${\cal H}_b^q$
are then given by
\beq
|q,n\ra= \frac{a^{\dg k(n+q-\frac{1}{k^2})}}{\sqrt{ \left[k\left(n+q-\frac{1}{k^2}\right) \right]!}} |0\ra.\label{fock-state1}
\eeq
It is easy to show that $Q_0, Q_{\pm}$ and $C$ act on these states as follows
\beqa
Q_0|q,n\ra &=& (q+n)|q,n\ra, \n
Q_+|q,n\ra &=& \pd_{i=1}^{k}\left( n+q+ \frac{i k-1}{k^2}\right)^\frac{1}{2}|q,n+1\ra, \n
Q_-|q,n\ra &=& \pd_{i=1}^{k}\left( n+q -\frac{(i-1)k+1}{k^2}\right)^\frac{1}{2}|q,n-1\ra, \n
C|q,n \ra  &=& \pd_{i=1}^{k}\left(\frac{i-k}{k}- \frac{1}{k^2}\right)|q,n \ra, \n
n &=& 0,1,\cdots, \hspace{1cm} q=\frac{1}{k^2},~\frac{k+1}{k^2},~\cdots,~ \frac{(k-1)k+1}{k^2}.\label{su11-rep}
\eeqa



\sect{Polynomial algebras via Jordan-Schwinger type construction}

The unitary representations of the polynomial algebras discussed in the preceding section are all
infinite dimensional. In this section, we shall employ a Jordan-Schwinger like construction \cite{Zhedanov92,Kumar01},
to derive polynomial algebras that have finite dimensional unitary representations. Towards this end,
we consider two mutually commuting polynomial algebras introduced in the preceding section,
$\left\{ Q_+^{(1)}, Q_-^{(1)}, Q_0^{(1)} \right\}$ of degree $(k_1-1)$ and
$ \left\{Q_+^{(2)}, Q_-^{(2)}, Q_0^{(2)} \right\}$ of degree $(k_2-1)$, where $k_1,k_2 =1,2,\cdots$.

Introduce new generators,
\beq
{\cal{Q}}_+ = Q_+^{(1)}Q_-^{(2)}, ~~~~
{\cal{Q}}_- = Q_+^{(2)}Q_-^{(1)}, ~~~~
{\cal{Q}}_0 = \frac{1}{2}\left(  Q_0^{(1)}- Q_0^{(2)} \right). \label{jordan-schwinger}
\eeq
We can easily show that ${\cal Q}_{0,\pm}$ form a polynomial algebra of degree $(k_1+k_2-1)$ which
close under the following commutation relations:
\beqa
\left[{\cal{Q}}_0,{\cal{Q}}_{\pm}\right] &=& \pm {\cal Q}_{\pm}, \n
\left[{\cal{Q}}_+,{\cal{Q}}_- \right] &=& {\cal \varphi}^{(k_1+k_2)}({\cal{Q}}_0,{\cal{L}})
-{\cal \varphi}^{(k_1+k_2)}({\cal{Q}}_0-1,{\cal{L}}),\label{su2-poly}
\eeqa
where
\beq
{\cal L} =\frac{1}{2}\left(Q_0^{(1)}+ Q_0^{(2)} \right)
\eeq
is the central element of the algebra,
\beq
\left[ {\cal L} , {\cal {Q}}_{\pm,0}\right] = 0
\eeq
and
\beq
{\cal \varphi}^{(k_1+k_2)}({\cal{Q}}_0, {\cal L})= -\pd_{i=1}^{k_1}\left({\cal L}+{\cal{Q}}_0+\frac{i}{k_1}
- \frac{1}{k_1^2}\right)\pd_{j=1}^{k_2}\left({\cal L}-({\cal{Q}}_0+1)+\frac{j}{k_2}- \frac{1}{k_2^2}\right)
\eeq
is a $(k_1+k_2)^{th}$-order polynomial in $Q_0$ and the central elements $\cal L$. The Casimir operator
of (\ref{su2-poly}) is given by
\beq
{\cal C}={\cal{Q}}_-{\cal{Q}}_+ +\varphi^{(k_1+k_2)}({\cal{Q}}_0,{\cal L})={\cal{Q}}_+{\cal{Q}}_-
+ \varphi^{(k_1+k_2)}({\cal{Q}}_0-1,{\cal L}).
\eeq
For $k_1+k_2=2$, i.e. $k_1=k_2=1$, the polynomial algebra (\ref{su2-poly}) reduces to the linear $su(2)$ algebra.
So the algebras defined by (\ref{su2-poly}) are polynomial deformations of $su(2)$.

In terms of two sets of mutually commuting boson operators acting on
the tensor product of the Fock spaces, we have the realization ($i=1,2$ below)
\beqa
Q_+^{(i)} =\frac{1}{(\sqrt{k_i})^{k_i}}(a_2^{\dg})^{k_i} , \hspace{1cm} Q_-^{(i)} = \frac{1}{(\sqrt{k_i})^{k_i}}(a_2)^{k_i} ,
\hspace{1cm}  Q_0^{(i)} =\frac{1}{k_i} \left( N_i+ \frac{1}{k_i} \right). \label{su2-poly-boson}
\eeqa
This realization gives rise to finite dimensional representations of the polynomial algebra (\ref{su2-poly}).
To show this, let $|q_1,n_1 \ra, |q_2,n_2 \ra$ be the one-mode Fock states of the algebras $
\left\{Q_{0,\pm}^{(1)} \right\}, \left\{ Q_{0,\pm}^{(2)} \right \}$ respectively, where $n_1,~n_2=0,1,\cdots$, and
$ q_1= \frac{1}{k_1^2}, \frac{k_1+1}{k_1^2}, \cdots, \frac{(k_1-1)k_1+1}{k_1^2}$ and
$q_2= \frac{1}{k_2^2}, \frac{k_2+1}{k_2^2}, \cdots, \frac{(k_2-1)k_2+1}{k_2^2}$.
The representations of $\left\{ {\cal Q}_{0,\pm} \right\} $ corresponding to the realization
(\ref{su2-poly-boson}) are then given by the two-mode Fock states $|q_1,n_1\ra|q_2,n_2\ra$.
Since $\cal L$ is a central element of the algebra, it must be a constant, denoted as $l$ below,
on any irreducible representations. This imposes a constraint on the values of $n_1$ and $n_2$,
\beqa
{\cal L} |q_1,n_1\ra|q_2,n_2\ra &=& \frac{1}{2}(q_1+n_1+q_2+n_2)|q_1,n_1\ra|q_2,n_2\ra\n
 &=&l\, |q_1,n_1\ra|q_2,n_2\ra. 	
\eeqa
That is $n_1+n_2=2l-(q_1+q_2)$. Thus obviously $2l-q_1-q_2$ take only positive integer values, i.e.
\beq
2l-q_1-q_2=0,1, \cdots.
\eeq
It follows that the Fock states corresponding to the realization (\ref{su2-poly-boson}) are
\beqa
|q_1,q_2,n,l\ra &=&|q_1,n\ra |q_2,2l-q_1-q_2-n\ra \n
&=& \frac{(a_1^{\dg})^{k_1\left(n+q_1-\frac{1}{k_1^2} \right)}(a_2^{\dg})^{k_2\left(2l-q_1-n-\frac{1}{k_2^2} \right)}}
{\sqrt{\left( k_1(n+q_1-\frac{1}{k_1^2})\right)!}\sqrt {\left( k_2(2l-q_1-\frac{1}{k_2^2}-n) \right)!}}|0 \ra,\n
\n
n&=&0,1,\cdots,2l-q_1-q_2,
\eeqa
noting that  $2l-q_1-q_2$ is always less than or equal to $2l-q_1-\frac{1}{k_2^2}$.
This gives us the $2l-q_1-q_2+1$ dimensional irreducible representation of ({\ref{su2-poly}),
\beqa
{\cal Q}_0|q_1,q_2,n,l\ra &=& (q_1-l+n)|q_1,q_2,n,l\ra, \n
{\cal Q}_+|q_1,q_2,n,l\ra &=&\pd_{i=1}^{k_2}\left( 2l-q_1-n-\frac{k_2(i-1)+1}{k_2^2}  \right)^\frac{1}{2} \n
&& \times \pd_{j=1}^{k_1}\left( n+q_1+ \frac{jk_1-1}{k_1^2}\right)^\frac{1}{2}|q_1,q_2,n+1,l\ra,\n
{\cal Q}_-|q_1,q_2,n,l\ra &=&  \pd_{i=1}^{k_2}\left( 2l-q_1-n +\frac{ik_2-1}{k_2^2}\right)^\frac{1}{2} \n
&& \times \pd_{j=1}^{k_1}\left( n+q_1 -\frac{(j-1)k_1+1}{k_1^2}\right)^\frac{1}{2}|q_1,q_2,n-1,l\ra.
\eeqa

By using the Fock-Bargmann correspondence,
\beq
a_i^{\dg} \longrightarrow z_i , \hspace{1cm} a_i \longrightarrow \frac{d}{dz_i},
\hspace{1cm} |n_i\ra \longrightarrow \frac{z_i^{n_i}}{\sqrt{n_i!}},
\eeq
we can make the following association
\beqa
|q_1,q_2,n,l\ra  \longrightarrow \frac{z_1^{k_1(n+q_1-\frac{1}{k_1^2})}z_2^{k_2(2l-q_1-n-\frac{1}{k_2^2})}}
{\sqrt{\left( k_1(n+q_1-\frac{1}{k_1^2})\right)!}\sqrt{\left( k_2(2l-q_1-\frac{1}{k_2^2}-n)\right)!}}.
\eeqa
Now since $l,q_1,q_2,k_1,k_2$ are constants, we can map the states
$|q_1,q_2,n,l\ra$ above to the monomials in $z=z_1^{k_1}/z_2^{k_2}$,
\beqa
\Psi_{q_1,q_2,n,l}(z)&=&\frac{z^n}{\sqrt{\left( k_1(n+q_1-\frac{1}{k_1^2})\right)!}\sqrt{\left( k_2(2l-q_1-\frac{1}{k_2^2}-n)\right)!}},\n
\n
n&=&0,1,\cdots, 2l-q_1-q_2.
\eeqa
The corresponding single-variable differential operator realization of (\ref{su2-poly}) takes the following form
\beqa
{\cal Q}_0 &=& z\frac{d}{dz}+ q_1 -l, \n
{\cal Q}_+ &=& z\frac{(\sqrt{k_2})^{k_2}}{(\sqrt{k_1})^{k_1}}\pd_{j=1}^{k_2}\left( 2l-q_1-\frac{(j-1)k_2+1}{k_2^2}-z\dz  \right), \n
{\cal Q}_- &=& z^{-1}\frac{(\sqrt{k_1})^{k_1}}{(\sqrt{k_2})^{k_2}} \pd_{j=1}^{k_1}
   \left(z\dz+q_1- \frac{(j-1)k_1+1}{k_1^2} \right).\label{su2-poly-d}
\eeqa
These differential operators form the same $2l-q_1-q_2+1$ dimensional representations in the space of polynomials
as those realized by (\ref{su2-poly-boson}) in the corresponding Fock space. We remark that because $\pd_{j=1}^{k_1}
   \left(q_1- \frac{(j-1)k_1+1}{k_1^2} \right)\equiv 0$ for all the allowed $q_1$ values there is no $z^{-1}$ term
in ${\cal Q}_-$ above and thus the differential operator expressions (\ref{su2-poly-d}) are non-singular.


\sect{Exact solution of the 2-mode boson systems}

We now use the differential operator realization (\ref{su2-poly-d})
to exactly solve the 2-mode boson Hamiltonian (\ref{bosonH1}).

By means of the Jordan-Schwinger type construction (\ref{jordan-schwinger}) and the realization (\ref{su2-poly-boson}),
identifying $k_1$ with $s$ and $k_2$ with $r$, we may express the Hamiltonian (\ref{bosonH1}) in terms of the generators
of the polynomial algebra (\ref{su2-poly}),
\beq
H=\sum_i^2 w_iN_i+\sum_{i,j}^2 w_{ij}N_iN_j + g\sqrt{s^sr^r}\left( {\cal Q}_+ + {\cal Q}_- \right)\label{bosonH2}
\eeq
with the number operators having the following expressions in ${\cal Q}_0$ and $\cal L$
\beq
N_1=s({\cal Q}_0+{\cal L}) -\frac{1}{s},~~~~~N_2=r({\cal L}-{\cal Q}_0) -\frac{1}{r}.
\eeq
Keep in mind that $\{{\cal Q}_{\mp,0}\}$ in (\ref{bosonH2}) as realized by (\ref{su2-poly-boson}) (and (\ref{jordan-schwinger})) form
the $(2l-q_1-q_2)+1$ dimensional representation of the polynomial algebra (\ref{su2-poly}). This representation is also
realized by the differential operators (\ref{su2-poly-d}) acting on the $(2l-q_1-q_2)+1$  dimensional space of polynomials with basis
 $\left\{1,z,z^2,...,z^{2l-q_1-q_2} \right \} $. We can thus equivalently represent (\ref{bosonH2}) (i.e. (\ref{bosonH1}))
as the single-variable differential operator of order max$\{s,r, 2\}$,
\beqa
H &=&\sum_i^2 w_iN_i+\sum_{i,j}^2 w_{ij}N_iN_j\n
 & &+gz \pd_{j=1}^{r} r\left( 2l-q_1-\frac{(j-1)r+1}{r^2}-z\dz \right)\n
 & &+gz^{-1}\pd_{j=1}^{s} s\left(z\dz+q_1- \frac{(j-1)s+1}{s^2} \right)\label{differentialH}
\eeqa
with
\beq
N_1=s(z\frac{d}{dz}+ q_1 ) -\frac{1}{s},~~~~~
N_2=r(2l-q_1-z\frac{d}{dz}) -\frac{1}{r}.
\eeq

We will now solve for the Hamiltonian equation
\beq
H\psi(z)=E\,\psi(z)\label{hamilton-eqn}
\eeq
 by using the Functional Bethe Ansatz method,
where $\psi(z)$ is the eigenfunction and $E$ is the corresponding eigenvalue. It is easy to verify
\beq
Hz^m=z^{m+1}\,g\pd_{j=1}^{r} r\left( 2l-q_1-\frac{(j-1)r+1}{r^2}-m\right)
   +{\rm lower~order~terms},~~~~m\in {\bf Z}_+.\label{hzm=hzm+1}
\eeq
This means that the differential operator ({\ref{differentialH}) is not exactly solvable.
However, it is quasi exactly solvable, since it has an invariant polynomial subspace of degree $(2l-q_1-q_2)+1$:
\beq
H{\cal V} \subseteq  {\cal V}, ~~~~~{\cal V}= {\rm span} \{1,z,...,z^{2l-q_1-q_2}\},~~~~~{\rm dim}{\cal V}=2l-q_1-q_2+1.
\eeq
This is easily seen from the fact that when $m=2l-q_1-q_2$ the first term on the r.h.s. of (\ref{hzm=hzm+1}) becomes
$z^{2l-q_1-q_2+1}\,g\pd_{j=1}^{r} r\left( q_2-\frac{(j-1)r+1}{r^2}\right)$ which vanishes identically for all
the allowed $q_2$ values. We remark that the quasi-exact solvability of the system is connected with its quantum integrability, 
i.e. with the fact that there exists quantum operator coinciding with a
linear combination of the operators $N_1$ and $N_2$ which commutes with the Hamiltonian (\ref{bosonH1}). 

As (\ref{differentialH}) is a quasi exactly solvable differential operator preserving ${\cal V}$,
up to an overall factor, its eigenfunctions have the form,
\beq
\psi(z)= \prod_{i=1}^M\left(z-\alpha_i\right),\label{w-function}
\eeq
where $M\equiv 2l-q_1-q_2~(=0,1,\cdots)$, and $\{\alpha_i\,|\,i=1,2,\cdots,M\}$ are roots of the polynomial
which will be specified later by the associated Bethe ansatz equations (\ref{bethe-ansatz-eqn}) below.
We can rewrite the Hamiltonian (\ref{differentialH}) as
\beq
 H = \sum_{i=1}^{\textrm{\tiny{max}}\{ r,s,2 \}} P_i(z)\left(\frac{d}{dz} \right)^i + P_0(z)\label{expansionH}
\eeq
where
\beqa
P_0(z)&=&zg\pd_{i=1}^{r}r\left(2l-q_1-\frac{(i-1)r+1}{r^2} \right) \n
&&+w_{11}\left( s q_1- \frac{1}{s} \right)^2 + w_{22}\left(  r\left(2l-q_1 \right)-\frac{1}{r}  \right)^2 \n
&&+2w_{12}\left( s q_1- \frac{1}{s} \right)\left(  r\left(2l-q_1 \right)-\frac{1}{r}  \right) \n
&& +w_{1}\left( s q_1- \frac{1}{s} \right) + w_{2}\left(  r\left(2l-q_1 \right)-\frac{1}{r}  \right)
\eeqa
and $P_i(z)$ are the coefficients in front of $d^i/dz^i$ in the expansion of (\ref{differentialH}) (see the Appendix),
\beqa
P_i(z) &=& g\,s^s\,z^{i-1} \sum_{k=i}^{s}
\left(\sum_{l_1<...<l_{k}}^{s} \pd_{j\neq l_1\neq\cdots\neq l_k}^s\,A_j\right)L_{k,i} \n
&&+ g\,(-r)^r\,z^{i+1} \sum_{k=i}^{r}
\left(\sum_{l_1<...<l_{k}}^{r}\pd_{j\neq l_1\neq\cdots\neq l_k}^r\,B_j\right)L_{k,i} \n
&&+ F\delta_{i,2}z^2 + D\delta_{i,1}z.
\eeqa
In the above expression, 
\beqa
A_i &=& q_1 - \frac{(i-1)s+1}{s^2},\n
B_i &=&  - \left(2l-q_1-\frac{(i-1)r+1}{r^2} \right),\n
L_{k,k} &=& 1, \n
L_{k,i} &=& \sum_{n_1 <...<n_{k-i}}^{k-1}n_1(n_2-1)...(n_{k-i}- (k-i)+1), \hspace{.5cm} i<k, \n
F   &=& w_{22}r^2+w_{11}s^2- 2w_{12}sr, \n
D   &=& w_{{22}}r^2 \left( 1- 2(2l-q_1-\frac{1}{r^2})\right) +w_{{11}} s^2\left(1+2(q_1- \frac{1}{s^2}) \right)\n
 &&+ 2w_{{12}} rs \left( 2(l-q_{{1}})+{\frac {1}{s^2}}-{\frac {1}{r^2}}-1  \right)+w_{{1}}s-w_{{2}}r. 
\eeqa

Dividing the Hamiltonian equation $H\psi= E\psi $ over by $\psi$ gives us
\beq
E= \frac{H\psi}{\psi} = \sum_{i=1}^{\textrm{\tiny{max}}\{ r,s,2 \}}P_i(z)i!\sum_{l_1<l_2<...<l_i}^M
\frac{1}{(z-\alpha_{l_1})...(z-\alpha_{l_i})}  +P_0(z).\label{e=hpsi/psi}
\eeq
The l.h.s. of (\ref{e=hpsi/psi}) is a constant, while the r.h.s is a meromorphic function in $z$ with at most simple poles.
For them to be equal, we need to eliminate all singularities on the r.h.s of (\ref{e=hpsi/psi}). We may achieve this by demanding
that the residues of the simple poles, $z=\alpha_i, i=1,2,..., M$ should all vanish. This leads to the Bethe ansatz
equations for the roots $\{\alpha_i \}$ :
\beq
\sum_{i=2}^{\textrm{\tiny{max}}\{ r,s,2 \}}\sum_{l_1<l_2<...<l_{i-1} \ne p}^M\frac{P_i(\alpha_p)i!}{(\alpha_p-\alpha_{l_1})\cdots
(\alpha_p-\alpha_{l_{i-1}})}+P_1(\alpha_{p})=0,~~~~~p=1,2,\cdots,M.   \label{bethe-ansatz-eqn}
\eeq
The wavefunction $\psi(z)$ (\ref{w-function}) becomes the eigenfunction of $H$ (\ref{differentialH}) in the space ${\cal V}$
provided that the roots $\{\alpha_i\}$ of the polynomial $\psi(z)$ (\ref{w-function}) are the solutions of (\ref{bethe-ansatz-eqn}).

Some remarks are in order. It is easily seen (from (\ref{w-function}) and (\ref{q-action}) below) that 
$H\psi/\psi$ is regular at $z=\pm\infty$. When (\ref{bethe-ansatz-eqn}) is satisfied, the r.h.s. of (\ref{e=hpsi/psi}) is 
analytic everywhere in the whole complex plane and thus must be a constant by the Liouville theorem.
Therefore the Bethe ansatz equation (\ref{bethe-ansatz-eqn}) is not only necessary but also sufficient condition
for the r.h.s. of (\ref{e=hpsi/psi}) to be independent of $z$.

To get the corresponding eigenvalue $E$, we consider the leading order expansion of $\psi(z)$,	
$$\psi(z)= z^M- z^{M-1}\sum_{i=1}^M\alpha_{i} +\cdots. $$
It is easy to show that ${\cal Q}_{\pm, 0}\psi(z)$ have the  expansions,
\beqa
{\cal Q}_+\psi &=& 
  -z^M \frac{(\sqrt{r})^{r}}{(\sqrt{s})^{s}}\left[\pd_{j=1}^{r}\left( q_2+1-\frac{(j-1)r+1}{r^2} \right)\right]
  \sum_{i=1}^M\alpha_i +\cdots,  \n
{\cal Q}_-\psi &=& z^{M-1}\frac{(\sqrt{s})^{s}}{(\sqrt{r})^{r}} \pd_{j=1}^{s}\left(2l-q_2- \frac{(j-1)s+1}{s^2} \right)+ \cdots,  \n
{\cal Q}_0\psi &=& z^{M}(l-q_2 )+  \cdots. \label{q-action}
\eeqa
Substituting these expressions into the Hamiltonian equation (\ref{hamilton-eqn}) and equating the $z^M$ terms, we arrive at
\beqa
E &=& w_{11}\left( s(2l- q_2 )- \frac{1}{s}\right)^2+w_{22}\left( rq_2 -\frac{1}{r} \right)^2\n
&&+ 2w_{12}\left( s(2l- q_2 )- \frac{1}{s}\right)\left( rq_2  -\frac{1}{r} \right)  \n
&&+w_{1}\left(s(2l- q_2) -\frac{1}{s}  \right)+w_2\left(rq_2  -\frac{1}{r} \right)  \n
&&-g \left[\pd_{j=1}^{r} r\left(  q_2+1-\frac{(j-1)r+1}{r^2}\right)\right]
  \sum_{i=1}^M\alpha_i ,\label{energy-generalH}
\eeqa
where $\{\alpha_i\}$ satisfy the Bethe ansatz equations (\ref{bethe-ansatz-eqn}). This gives
the eigenvalue of the 2-mode boson Hamiltonian (\ref{bosonH1}) with the corresponding eigenfunction $\psi(z)$ (\ref{w-function}).

\sect{ Explicit examples corresponding to BECs}

We will now work out in complete detail the Bethe ansatz equations and energy eigenvalues
of the Hamiltonian (\ref{bosonH1})
for the special cases of $s,r \le 2$ and $r=s=3$. These models arise in the description of Josephson tunneling effects and
atom-molecule conversion processes in the context of BECs. \\

\noindent \textbf{A. $s=1, r=1$}\\

The Hamiltonian is
\beq
H=\sum_i^2 w_iN_i+\sum_{i,j}^2 w_{ij}N_iN_j +  g\left( a_1^{\dagger }a_2 +a_1a_2^{\dagger } \right).\label{h11}
\eeq
This is the so-called two coupled BEC model and has been solved in \cite{Links03}
via a different method, i.e. the ABA method. From the general results in the preceding section,
in this case, we have $q_1=q_2=1$, which means that $2l-q_1-q_2=2(l-1)=0,1,\cdots$.
That is $l-1=0,\frac{1}{2},1,\cdots$. The differential operator representation of the Hamiltonian
(\ref{h11}) is
\beq
H= P_2(z) \frac{d^2}{dz^2}+ P_1(z) \dz +P_0(z),
\eeq
where
\beqa
P_2(z)&=&A_{11}z^2,\n
P_1(z)&=&-gz^2+B_{11}z+g,\n
P_0(z)&=&2(l-1)g z+D_{11}
\eeqa
with
\beqa
A_{11} &=& w_{{11}}+w_{{22}}-2w_{{12}}\neq 0, \n
B_{11} &=& w_{{1}}-w_{{2}}+w_{{11}} + \left( 5-4l\right)w_{{22}}
  + \left( 4l-6\right) w_{{12}},\n
D_{11} &=& 2 (l-1)w_2+4(l-1)^2w_{22}.
\eeqa
The Bethe ansatz equations are given by
\beqa
\sum_{i \ne p}^{2(l-1)} \frac{2}{\alpha_i-\alpha_p}
=\frac{g+B_{11}\alpha_p-g\alpha_p^2}{A_{11}\alpha_p^2},~~~~~p=1,2,\cdots,2(l-1)
\eeqa
and the energy eigenvalues are
\beqa
E = 4w_{{11}} (l-1)^2+2w_1(l-1)-g \sum_{i=1}^{2(l-1)}\alpha_{{i}}.
\eeqa

\noindent\textbf{B. $s=2, r=1$}\\

The Hamiltonian is
\beq
H=\sum_i^2 w_iN_i+\sum_{i,j}^2 w_{ij}N_iN_j +  g\left( a_1^{\dagger 2}a_2 +a_1^2a_2^{\dagger } \right).\label{h21}
\eeq
This is the homo-atomic-molecular BEC model and has been solved by the ABA method \cite{Links03}.
Specializing the general results in the preceding section to this case,
we have $q_1=\frac{1}{4}$ or $\frac{3}{4}$, and $q_2=1$.
The differential operator representation of the Hamiltonian (\ref{h21}) is thus
\beq
H= P_2(z) \frac{d^2}{dz^2}+ P_1(z) \dz +P_0(z)
\eeq
where
\beqa
P_2(z)&=&A_{21}z^2+4gz,\n
P_1(z)&=&-gz^2+B_{21}z+8gq_1,\n
P_0(z)&=&g(2l-q_1-1)z+D_{21}
\eeqa
with
\beqa
A_{21} &=&  4\,w_{{11}}+w_{{22}}-4\,w_{{12}}, \n
B_{21} &=&  2\,w_{{1}}-w_{{2}}+2w_{{11}} \left( 1+4\,q_{{1
}} \right) +w_{{22}} \left( 3+2\,q_{{1}}-4\,l \right) \n
& & + w_{{12}}  \left( -7-8\,q_{{1}}+8\,l \right), \n
D_{21} &=& 2w_{{1}} \left( q_{{1}}-\frac{1}{4} \right)+w_{{2}} \left( 2\,l-q_{{1}}-1 \right)\n
& &+ 4w_{11}\left(q_1-\frac{1}{4}\right)^2+w_{22}(2l-1-q_1)^2\n
& &+4 w_{{12}}  \left( q_1-\frac{1}{4}\right)(2l-1-q_1).
\eeqa
The Bethe ansatz equations are
\beqa
\sum_{i \ne p}^{2l-1-q_1} \frac{2}{\alpha_i-\alpha_p}=\frac{8gq_1+B_{21}\alpha_p-g\alpha_p^2}
  {\alpha_p(A_{21}\alpha_p+4g)},~~~~~p=1, 2, \cdots, 2l-1-q_1
\eeqa
and the energy eigenvalues are given by
\beqa
E = 2w_1\left(2l-\frac{5}{4}\right)+ 4w_{{11}} \left( 2l-\frac{5}{4}\right) ^{2}- g \sum_{i=1}^{2l-1-q_1}\alpha_i .
\eeqa

\noindent \textbf{C. $s=2, r=2$}\\

The Hamiltonian is
\beq
H=\sum_i^2 w_iN_i+\sum_{i,j}^2 w_{ij}N_iN_j +  g\left( a_1^{\dagger 2}a_2^2 +a_1^2a_2^{\dagger 2 } \right).\label{h22}
\eeq
This gives another model of the atom-molecule BECs. To our knowledge, this model has not been exactly solved previously.
Applying the general results in the preceding section, we have in this case
$q_1 = \frac{1}{4},~\frac{3}{4}$ and $q_2 = \frac{1}{4},~ \frac{3}{4}$.
The differential operator representation of the Hamiltonian (\ref{h22}) is
\beq
H= P_2(z) \frac{d^2}{dz^2}+ P_1(z) \dz +P_0(z),
\eeq
where
\beqa
P_2(z)&=&4gz^3+4A_{22}z^2+4gz,\n
P_1(z)&=&B_{22}z^2+D_{22}z+8gq_1,\n
P_0(z)&=&F_{22}z+G_{22},
\eeqa
with
\beqa
A_{22} &=& w_{{11}}+w_{{22}}-2 w_{{12}},\n
B_{22} &=&   8\,g ( 1+q_{{1}} -2l) ,\n
D_{22}&=&2\,w_{{1}}-2\,w_{{2}}
  +2w_{{11}} \left( 1+4\,q_{{1}} \right) +2w_{{22}} \left( 3-8\,l+4\,q_{{1}} \right) \n
&& +8 w_{{12}}  \left( -1-2\,q_{{1}}+2\,l \right), \n
F_{22} &=&  4g\left( 2\,l-q_{{1}} -\frac{1}{4}\right)  \left(2\,l-q_{{1}}-\frac{3}{4} \right),\n
G_{22}&=&2w_{{1}} \left( q_{{1}}-\frac{1}{4} \right)+2w_{{2}} \left( 2\,l-\,q_{{1}}-\frac{1}{4} \right)
  +4w_{{11}} \left( q_1-\frac{1}{4} \right)^2  \n
&& + 8 w_{{12}} \left(q_1-\frac{1}{4}\right)\left(2l-q_1-\frac{1}{4}\right)
  +4w_{22}\left(2l-q_1-\frac{1}{4}\right)^2.
\eeqa
Note that $(2l-q_1-q_2)(2l-q_1+q_2-1)\equiv (2l-q_1-1/4)(2l-q_1-3/4)$ for $q_2=1/4, 3/4$.
The Bethe ansatz equations read
\beqa
\sum_{i \ne p}^{2l-q_1-q_2} \frac{2}{\alpha_i-\alpha_p}=\frac{8gq_1+D_{22}\alpha_p-B_{22}\alpha_p^2}
{4\alpha_p(g\alpha_p^2+A_{22}\alpha_p+g)},~~~~~p=1,2,\cdots,2l-q_1-q_2
\eeqa
and the energy eigenvalues are
\beqa
E&=&4w_{{11}} \left( 2l-q_2-\frac{1}{4} \right) ^{2}+4w_{{22}} \left( q_2-\frac{1}{4}\right)^2\n
&&+8 w_{{12}}\left(2l-q_2-\frac{1}{4}\right)\left(q_2-\frac{1}{4}\right)
 +2w_{{1}} \left( 2l-q_2-\frac{1}{4}\right) \n
&& +2w_{{2}} \left( q_2-\frac{1}{4}\right) -4\,g \left( q_{{2}}+\frac{1}{4}\right)
 \left( q_{{2}}+\frac{3}{4}\right)  \sum_{i=1}^{2l-q_1-q_2}\alpha_{{i}}.
\eeqa

\noindent \textbf{D. $s=3, r=3$}\\

The considered examples with $s, r\leq 2$ may in principle be treated using the ABA method
based on Lie algebra su(2) (without any polynomial deformations). 
We now present an explicit example for which the ABA method is not applicable . The Hamiltonian is 
\beq
H=\sum_i^2 w_iN_i+\sum_{i,j}^2 w_{ij}N_iN_j +  
g\left( a_1^{\dagger 3}a_2^3 +a_1^3a_2^{\dagger 3 }. \right).\label{h33}
\eeq
This is a non-linear optical model with third-order harmonic generation. 
Specializing the general results in the preceding section to this case,
we have $ q_1, q_2 =\frac{1}{9} ,\frac{4}{9}$, or $\frac{7}{9}$.  
The differential operator representation of the Hamiltonian (\ref{h33}) is
\beq
H= P_3(z) \frac{d^3}{dz^3} +P_2(z)\frac{d^2}{dz^2}+ P_1(z) \dz +P_0(z),
\eeq
where
\beqa
P_3(z) &=& 27g(-z^4+ z^2 ) \n 
P_2(z) &=& A_{33}z^3 +B_{33}z^2 +D_{33}z,\n
P_1(z) &=& F_{33}z^2+G_{33}z+K_{33} \n 
P_0(z)&=&R_{33}z +S_{33}
\eeqa
with 
\beqa 
A_{33} &=& 9g \left(18 l-9q_1-13\right), \n 
B_{33} &=& 9 (w_{11}+w_{22}-2w_{12}), \n
D_{33} &=& 9g \left( 9q_1+5\right), \n 
F_{33} &=& 9g\left(-36l^2-\frac{76}{9} +34 l + 36 lq_1 -9q_1^2-17 q_1\right), \n 
G_{33} &=& 3w_1-3w_2 +w_{11}(7+18q_1)+2w_{12}(-9-18q_1+18l) +w_{22}(7+18q_1),  \n 
K_{33} &=& 9g\left( q_1+9q_1^2+\frac{4}{9} \right), \n 
R_{33} &=& 27g\left(2l-q_1-\frac{1}{9} \right)\left(2l-q_1-\frac{4}{9} \right)\left(2l-q_1-\frac{7}{9} \right), \n 
S_{33} &=& 9w_{11}\left(  q_1- \frac{1}{9} \right)^2 + 9w_{22}\left(  2l-q_1-\frac{1}{9}  \right)^2 +18 w_{12}
   \left( q_1- \frac{1}{9} \right)\left( 2l-q_1 -\frac{1}{9}  \right) \n
&& +3w_{1}\left( q_1- \frac{1}{9} \right) + 3w_{2}\left(  2l-q_1 -\frac{1}{9}  \right).
\eeqa 
The Bethe ansatz equations read
\beqa
&&\sum_{i<j \ne p}^{2l-q_1-q_2}\frac{162g(\alpha_p^4-\alpha_p^2)}{(\alpha_i-\alpha_p)(\alpha_j-\alpha_p)} 
+ \sum_{i \ne p}^{2l-q_1-q_2}\frac{2(A_{33}\alpha_p^3+B_{33}\alpha_p^2+D_{33}\alpha_p)}{\alpha_i-\alpha_p} \n 
\n
&&~~~~~~~ = F_{33}\alpha_p^2+G_{33}\alpha_p+K_{33},~~~~~p=1,2,\cdots,2l-q_1-q_2
\eeqa
and the energy eigenvalues are
\beqa
E&=&9w_{{11}} \left( 2l-q_2-\frac{1}{9} \right) ^{2}+9w_{{22}} \left( q_2-\frac{1}{9}\right)^2\n
&&+18 w_{{12}}\left(2l-q_2-\frac{1}{9}\right)\left(q_2-\frac{1}{9}\right)
 +3w_{{1}} \left( 2l-q_2-\frac{1}{9}\right) \n
&& +3w_{{2}} \left( q_2-\frac{1}{9}\right) -27\,g \left( q_{{2}}+\frac{2}{9}\right)
 \left( q_{{2}}+\frac{5}{9}\right) \left( q_{{2}}+\frac{8}{9}\right)   \sum_{i=1}^{2l-q_1-q_2}\alpha_{{i}}.
\eeqa

\sect{Spectral Equivalence with QES Schr\"odinger Potentials}

The Hamiltonians in section 5 correspond to second order differential operators and can be mapped to
Schr\"odinger equations with QES potentials via a suitable similarity transformation and change of variables
\cite{Zaslavskii90}.

Explicitly, if $H$ is written in the following form
\beq
H=P(z)\frac{d^2}{dz^2} +\left(Q(z)+\frac{1}{2}P'(z)\right)\dz +R(z)
\eeq
then it can be mapped to a Schr\"odinger operator,
\beq
\tilde{H} = -e^{W(x)}H e^{-W(x)} = -\frac{d^2}{dx^2} + V(x),
\eeq
where the variable $x$ and $z$ are related by  
(we assume $z=z(x)$ is invertible on a certain interval to give $x=x(z)$ ) \cite{Ullate05},
\beq
x = x(z)=\pm \int^z \frac{dy}{\sqrt{P(y)}} \label{change-of-variable}
\eeq
and $W(x)$ is given as
\beq
W (x) =  \int^{z(x)} \frac{Q(y)}{2P(y)}dy .
\eeq
The potential function is given by
\beq
V(x)= \left.\left\{ -R(z)+\frac{1}{2}Q'(z)-\frac{Q(z)(P'(z)-Q(z))}{4P(z)} \right\}\right|_{z=z(x)}
\eeq
Then the solutions of the 2nd order ODE $H\psi(z)=E\psi(z)$ with
eigenvalue $E$ is mapped to solutions of the Schr\"odinger
equation
\beq
\tilde{H}\tilde{\psi}(x)=\tilde{E}\tilde{\psi}(x)
\eeq
with eigenvalue $\tilde{E}=-E$ and corresponding Schr\"odinger wavefunction
\beq
\tilde{\psi}(x)=e^{-W(x)}\,\psi(z(x)).
\eeq

We will not discuss the square integrability of the Schr\"odinger wavefunction $\tilde{\psi}(x)$, but
derive the explicit Schr\"odinger potentials corresponding to the special models in the preceding section. \\

\noindent {\bf I.} $s=1,r=1$:
\vskip.1in
For this case we have
\beqa
P(z)&=&A_{11}z^2,\n
Q(z)&=&-gz^2+(B_{11}-A_{11})z+g,\n
R(z)&=&2(l-1)g z+D_{11}.
\eeqa
{}From (\ref{change-of-variable}), we obtain
\beq
z(x)=e^{\sqrt{A_{11}}x}
\eeq
The potential is
\beqa
V(x) &=& \frac { \left( gz^{2}+(A_{11}-B_{11})z-g\right)
\left( gz^{2}+(3A_{11}-B_{11})z-g \right) }{4A_{11}z^{2}}\n
 && -g(2l-1)z+\frac{B_{11}-A_{11}}{2}-D_{11}\n
&=&\frac{g^2}{2A_{11}}\cosh(2\sqrt{A_{11}}x)+g\left(2-\frac{B_{11}}{A_{11}}\right)\sinh(\sqrt{A_{11}}x)\n
&&-(2l-1)ge^{\sqrt{A_{11}}x}-D_{11}+\frac{(A_{11}-B_{11})^2-2g^2}{4A_{11}}.
\eeqa

\noindent {\bf II.} $s=2, r=1$:
\vskip.1in
In this case we have,
\beqa
P(z)&=&A_{21}z^2+4gz,\n
Q(z)&=&-gz^2+(B_{21}-A_{21})z+2g(4q_1-1),\n
R(z)&=&g(2l-q_1-1)z+D_{21}
\eeqa
and $q_1=\frac{1}{4}$ or $\frac{3}{4}$. From (\ref{change-of-variable}), we derive
\beq
z(x) = \frac{2g}{A_{21}} \left(\textrm{cosh}( \sqrt{A_{21}}x)-1\right).\label{change2}
\eeq
The potential is
\beqa
V(x) &=& \frac { \left( g{z}^{2}+(A_{21}-B_{21})z +2g-8gq_1\right)  \left(g{z}^{2}+
(3A_{21}-B_{21})z+6g-8gq_1 \right) }{4(A_{21}{z}^{2}+4gz)}\n
& &-g(2l-q_1)z+ \frac{B_{21}-A_{21}}{2}-D_{21}\n
&=&\frac{g^2}{A_{21}^2}\tanh^2\left(\frac{\sqrt{A_{21}}}{2}x\right)\sinh^2\left(\frac{\sqrt{A_{21}}}{2}x\right)
\left[\frac{4g^2}{A_{21}}\sinh^2\left(\frac{\sqrt{A_{21}}}{2}x\right)+2A_{21}-B_{21}\right]\n
& &+\frac{1}{4A_{21}}\left[(A_{21}-B_{21})(3A_{21}-B_{21})
  +8g^2(1-2q_1)\right]\tanh^2\left(\frac{\sqrt{A_{21}}}{2}x\right)\n
& &-\frac{4(2l-q_1)g^2}{A_{21}}\sinh^2\left(\frac{\sqrt{A_{21}}}{2}x\right)\n
& &+\frac{(3-8q_1)A_{21}+(4q_1-2)B_{21}}{4\cosh^2\left(\frac{\sqrt{A_{21}}}{2}x\right)}
  +\frac{B_{21}-A_{21}}{2}-D_{21}.\label{potential21}
\eeqa
Here we have used  $(1-4q_1)(3-4q_1)=0$ for the two allowed $q_1$ values $q_1=\frac{1}{4}$ or $\frac{3}{4}$.

Let us consider the special case of $A_{21}=0$. In this case,
\beq
z(x)=gx^2
\eeq
as can be seen from the $A_{21}\rightarrow 0$ limit of (\ref{change2}).  The Schr\"odinger potential
(\ref{potential21}) then becomes
\beq
V(x)=\frac{g^4}{16}x^6-\frac{g^2}{8}B_{21}x^4+\frac{B_{21}^2+8g^2(1-4l)}{16}x^2+q_1B_{21}-D_{21}.
\eeq
This is a non-singular sextic potential. \\

\noindent {\bf III.} $s=2, r=2$:
\vskip.1in
For this case we have
\beqa
P(z)&=&4gz^3+4A_{22}z^2+4gz,\n
Q(z)&=&B_{22}z^2+(D_{22}-4A_{22})z+8gq_1-2g,\n
R(z)&=&F_{22}z+G_{22}
\eeqa
and $q_1=\frac{1}{4}$ or $\frac{3}{4}$. From (\ref{change-of-variable}), we obtain
\beq
z(x)=g^{-\frac{1}{3}}\,\wp(g^{\frac{1}{3}}x;g_2,g_3)-\frac{A_{22}}{3g},\label{change3}
\eeq
where $\wp(x;g_2,g_3)$ is Weierstrass's elliptic function with invariants $g_2$ and $g_3$ given by
\beq
g_2=\frac{4}{3}g^{\frac{2}{3}}\left(\frac{A_{22}^2}{g^2}-3\right),~~~~
g_3=\frac{4}{27}A_{22}\left(9 - \frac{2A_{22}^2}{g^2}\right).
\eeq
Hereafter we will denote $\wp(x;g_2,g_3)$ simply as $\wp(x)$.
The potential is computed as follows
\beqa
V(x) &=&  \left( B_{22}{z}^{2}+(D_{22}-4A_{22})z+8gq_1-2g\right) \n
&&\times \frac {(B_{22}-12g){z}^{2}+(D_{22}-12A_{22})z+8gq_1-6g}{16\left( g{z}^{3}+A_{22}z^2+gz\right)} \n
& & +(B_{22}-F_{22})z+\frac{D_{22}-4A_{22}}{2}-G_{22}\n
&=&\sum_{i=1}^4\,c_i\frac{\left(g^{-\frac{1}{3}}\wp(g^{\frac{1}{3}}x)-\frac{A_{22}}{3g}\right)^i}
  {4\wp'(g^{\frac{1}{3}}x)^2}+(B_{22}-F_{22})g^{-\frac{1}{3}}\wp(g^{\frac{1}{3}}x)\n
& &+\frac{A_{22}(B_{22}-F_{22})}{3g}+\frac{D_{22}-4A_{22}}{2}-G_{22},
\eeqa
where
\beqa
c_1&=&2g(4q_1-3)(D_{22}-4A_{22})+2g(4q_1-1)(D_{22}-12A_{22}),\n
c_2&=& 2g(4q_1-3)B_{22}+2g(4q_1-1)(B_{22}-12g)+(D_{22}-4A_{22})(D_{22}-12A_{22}),\n
c_3&=&B_{22}(D_{22}-12A_{22})+(B_{22}-12g)(D_{22}-4A_{22}),\n
c_4&=&B_{22}(B_{22}-12g).
\eeqa
Here we have used $4(gz^3+A_{22}z^2+gz)=4\wp(g^{\frac{1}{3}}x)^3-g_2\wp(g^{\frac{1}{3}}x)-g_3=\wp'(g^{\frac{1}{3}}x)^2$
and $(1-4q_1)(3-4q_1)=0$ for the two allowed $q_1$ values $q_1=\frac{1}{4}$ or $\frac{3}{4}$.

\sect{Discussion}

Let us now quickly summarize the work. We began by constructing the boson
representation of a class of $su(1,1)$ polynomially deformed algebras (\ref{su11-poly-alg}),  deriving their
infinite dimensional Fock space realization and lowest weight state parametrization. We then used
the Jordan-Schwinger like construction to
get the polynomial algebra (\ref{su2-poly}) which possesses finite dimensional irreducible representations.
We used the differential realization of (\ref{su2-poly}) to rewrite the Hamiltonian (\ref{bosonH1})
as QES differential operators acting on the finite dimensional monomial space.
The exact eigenfunctions and eigenvalues of the Hamiltonian were
then found by employing the Functional Bethe Ansatz technique.
As examples, we provided some explicit expressions for the BEC models which correspond to the $r,s \le 2$ cases of
(\ref{bosonH1}) and established the spectral correspondence of these specific models with QES Schr\"odinger potentials.

In deriving our results, we showed that in general the Hamiltonians defined in (\ref{bosonH1})
are QES differential operator of order 3 or higher. This paper provides an algebraization of 
such higher order QES differential operators and unravels the dynamical polynomial algebra symmetry of (\ref{bosonH1}).  
It also shows that the Functional Bethe Ansatz method provides a simple way to find exact eigenvalues and
eigenfunctions of such higher order differential operators.

There are a number of extensions that we plan to pursue in this line of investigation.
Firstly, we note that the Jordan-Schwinger like construction can be extended straightforwardly
to study other non-linear quantum optical models such as the general multi-mode boson Hamiltonians
of the form
\beqa
H&=&\sum_i^{k+k'}w_iN_i+ \sum_{i,j}^{k+k'} w_{ij}N_iN_j\n
& & + g\left( a_1^{\dagger m_1}\cdots a_k^{\dagger m_k}
a_{k+1}^{m_{k+1}}\cdots a_{k+k'}^{m_{k+k'}} + a_{1}^{m_{1}}\cdots a_{k}^{m_{k}}a_{k+1}^{\dagger m_{k+1}}
\cdots a_{k+k'}^{\dagger m_{k+k'}} \right).
\eeqa
Results on on this and other models of physical interest will be presented elsewhere.

\vskip.2in
\noindent {\bf Acknowledgments:} This work was supported by the Australian Research Council.
The authors would like to thank Ryu Sasaki for very valuable comments and suggestions which lead to
significant improvement of the presentation of the paper.

\sect{ Appendix}
In this appendix, we work out the expansion coefficients in front of $\frac{d^i}{dz^i}$ in the xpansion
of $\pd_{i=1}^{m} \left( z\dz +A_i \right)$.

First, we see
\beqa
\pd_{i=1}^{m} \left( z\dz +A_i \right) &=& \pd_{i=1}^{m} A_i + \left(\sum_{j_1=1}^m\pd_{i \ne j_1}^mA_i\right)z\dz
+ \left(\sum_{j_1 <j_2 }^m \pd_{i \ne j_1 \ne j_2}^m A_i \right)\left( z\dz \right)^2 \n
&& +\cdots + \left(\sum_{j_1 <j_2 <.. <j_m}^m  \pd_{i \ne j_1...\ne j_m}^m A_i \right)\left(z\dz \right)^m\n
\eeqa
and since
\beqa
\left(z\dz \right)^k &=& z^k \frac{d^k}{dz^k} + \left(\sum_{n=1}^{k-1}n \right)z^{k-1}\frac{d^{k-1}}{dz^{k-1}}
+ \left(\sum_{n_1 <n_2 }^{k-1}n_1(n_2-1) \right)z^{k-2}\frac{d^{k-2}}{dz^{k-2}} + \cdots\n
&=& \sum_{i=1}^kL_{k,i}z^i\frac{d^{i}}{dz^{i}}
\eeqa
where
\beqa
L_{k,k} &=& 1 \n
L_{k,i} &=& \sum_{n_1 <...<n_{k-i}}^{k-1}n_1(n_2-1)...(n_{k-i}- (k-i)+1), \hspace{.5cm} i<k
\eeqa
We can regroup the equation (8.1) as
\beqa
&&\pd_{i=1}^{m} \left( z\dz +A_i \right) \n
&&~~~~ =\pd_{i=1}^{m} A_i+ \left(L_{1,1}\left(\sum_{j_1=1}^m\pd_{i \ne j_1}^mA_i\right)
+\cdots+   L_{m,1} \left(\sum_{j_1 <j_2 <\cdots <j_m}^m  \pd_{i \ne j_1\cdots \ne j_m}^m A_i \right)\right) z\dz \n
&&~~~~~+\left(L_{2,2}\left(\sum_{j_1 <j_2 }^m \pd_{i \ne j_1 \ne j_2}^m A_i \right)+\cdots
+L_{m,2} \left(\sum_{j_1 <j_2 <\cdots <j_m}^m  \pd_{i \ne j_1\cdots\ne j_m}^m A_i \right) \right)z^2\frac{d^2}{dz^2}\n
&&~~~~~ +\textrm{higher order terms}\n
&=&\pd_{i=1}^{m} A_i +\sum_{i=1}^m \sum_{k=i}^{m}\left(\sum_{l_1<...<l_{k}}^{m}
\pd_{j\neq l_1\neq\cdots\neq l_k}^mA_j\right)L_{k,i}\,z^{i}\frac{d^i}{dz^i}.
\eeqa

\bebb{99}

\bbit{Higgs79}
P.W. Higgs, J. Phys. A: Math. Gen. {\bf 12}, 309 (1979).

\bbit{Rocek91}
M. Rocek, Phys. Lett. B {\bf 255}, 554 (1991).

\bbit{Schoutens91}
K Schoutens, A. Sevrin and P. Van Nieuwenhuizen, Comm. Math. Phys. {\bf 124}, 87 (1991);
  Phys. Lett. B {bf 255}, 549 (1991).

\bbit{Granovsky92}
Ya.I. Granovsky, A.S. Zhedanov and I.M. Lutzenko, Ann. Phys. (NY) {\bf 217}, 1 (1992).

\bbit{Letourneau95}
P. Letourneau and L. Vinet, Ann. Phys. (NY) {\bf 243}, 144 (1995).

\bbit{Quesne94}
C. Quesne, Phys. Lett. A {\bf 193}, 249 (1994); SIGMA {\bf 3}, Paper 067 (2007).

\bbit{Bonatsos94}
D. Bonatsos, C. Daskaloyannis and K. Kokkotas, Phys. Rev. A {\bf 50}, 3700 (1994).

\bbit{Karassiov94}
V.P. Karassiov and A. Klimov, Phys. Lett. A {\bf 191}, 117 (1994).


\bbit{Karassiov02}
V.P. Karassiov, A.A. Gusev and S.I. Vinitsky, Phys. Lett. A {\bf 295}, 247 (2002).

\bbit{Klishevich01} S.M. Klishevich and M.S. Plyushchay, Nucl. Phys. B {\bf 606}, 583 (2001);
 ibid {\bf 616}, 403 (2001).

\bbit{Smith90}
S.P. Smith, Trans. Amer. Math. Soc. {\bf 322}, 285 (1990).

\bbit{Zhedanov92}
A.S. Zhedanov, Mod. Phys. Lett. A {\bf 7}, 507 (1992).

\bbit{Kumar01}
V.S. Kumar, B.A. Bambah, and R. Jagannathan, J. Phys. A: Math. Gen. {\bf 38}, 34 (2001);
Mod. Phys. Lett. A {\bf 17}, 1559 (2002).



\bbit{Beckers99}
J. Beckers, Y. Brihaye, and N. Debergh, J. Phys. A: Math. Gen. {\bf 32}, 2791 (1999).
\bbit{Debergh00}
N. Debergh, J. Phys. A: Math. Gen. {\bf 33}, 7109 (2000).

\bbit{Turbiner88}
A. Turbiner, Comm. Math. Phys. {\bf 118}, 467 (1988);
Quasi-exactly-solvable differential equations, 1994, hep-th/9409068.

\bbit{Ushveridze94}
A.G. Ushveridze, Quasi-exactly solvable models in quantum mechanics, Institute of Physics
Publishing, Bristol, 1994.

\bbit{Gonzarez93}
A. Gonz\'arez-L\'opez, N. Kamran and P. Olver, Comm. Math. Phys. {\bf 153}, 117 (1993).

\bbit{Wiegmann94}
P.B. Wiegmann and A.V. Zabrodin, Phys. Rev. Lett. {\bf 72}, 1890 (1994);
 Nucl. Phys.  B {\bf 451}, 699 (1995).

\bbit{Sasaki09}
R. Sasaki, W.-L. Yang, and Y.-Z. Zhang, SIGMA {\bf 5}, Paper 104 (2009).





\bbit{Anderson95}
M.H. Anderson, J.R. Ensher, M.R. Matthews, C.E. Wieman and E.A. Cornell, Science {\bf 269}, 198 (1995).

\bbit{Anglin02}
J.R. Anglin and W. Ketterle, Nature {\bf 416}, 211 (2002).

\bbit{Zoller02}
P. Zoller, Nature {\bf 417}, 493 (2002).

\bbit{Donley02}
E.A. Donley, N.R. Claussen, S.T. Thompson and C.E. Wieman, Nature {\bf 417}, 529 (2002).

\bbit{Alvarez02} G. \'Alvarez, F. Finkel, A. Gonz\'alez-L\'opez and M.A. Rodr\'iguez,
  J. Phys. A: Math. Gen. {\bf 35}, 8705 (2002).
  
\bbit{Alvarez95} G. \'Alvarez and R.F. \'Alvarez-Estrada, J. Phys. A: Math. Gen. {\bf 28},
  5767 (1995); ibid {\bf 34}, 10045 (2001).

\bbit{Links03}
J. Links, H.-Q. Zhou, R.H. McKenzie, and M.D. Gould, J. Phys. A: Math. Gen.
{\bf 36}, R63 (2003).



\bbit{Zaslavskii90} O.B. Zaslavskii, Phys. Lett. A {\bf 149}, 365 (1990).

\bbit{Ullate05}
D. Gomez-Ullate, N. Kamran, and R. Milson, J. Phys. A: Math. Gen. {\bf 38}, 2005
(2005).


\eebb

\end{document}